\documentstyle[prd,aps,graphics]{revtex}

\begin{document}    
\draft

\twocolumn[\hsize\textwidth\columnwidth\hsize\csname
@twocolumnfalse\endcsname

\title{Damping scales of neutralino cold dark matter}

\author{Stefan Hofmann$^a$\footnotemark[1],
Dominik J. Schwarz$^{a,b}$\footnotemark[2],
Horst St\"ocker${}^a$}
\address{$^a$ Institut f\"ur Theoretische Physik,  
J.W.Goethe-Universit\"at,
Robert-Mayer-Str. 8-10, D-60054 Frankfurt a.M., Germany \\
$^b$ Institut f\"ur Theoretische Physik, 
Technische Universit\"at Wien,
Wiedner Hauptstra{\ss}e 8-10, A-1040 Wien, Austria}
\date{\today}

\maketitle

\begin{abstract}
The lightest supersymmetric particle, most likely the neutralino,
might account for a large fraction of dark matter in the Universe. We 
show that the primordial spectrum of density fluctuations in neutralino
cold dark matter (CDM) has a sharp cut-off due to two damping mechanisms:
collisional damping during the kinetic decoupling of the neutralinos 
at about $30$ MeV (for typical neutralino and sfermion masses) 
and free streaming after last scattering of neutralinos. The last 
scattering temperature is lower than the kinetic decoupling temperature 
by one order of magnitude. The cut-off in the primordial spectrum 
defines a minimal mass for CDM objects in hierarchical structure 
formation. For typical neutralino and sfermion masses the first 
gravitationally bound neutralino clouds have to have masses above 
$10^{-7} M_\odot$.
\end{abstract}
\vspace*{1cm}
]
\footnotetext[1]{\tt stehof@th.physik.uni-frankfurt.de} 
\footnotetext[2]{\tt dschwarz@hep.itp.tuwien.ac.at}

\section{Introduction}

Recent observations of the cosmic microwave background (CMB) anisotropies
are consistent with a key prediction of inflationary cosmology: The universe
appears to be spatially flat \cite{MAXIBOOM}. However, only a small percentage
of the mass that is needed to account for the critical energy density of the
universe comes in the form of baryons. Recent observations of primordial 
deuterium and other light elements suggest that the baryonic mass density is 
$\omega_{\rm b} = 0.019 \pm 0.0024 $ \cite{Tytler}, which implies that 
only about $5$ per cent of the mass in the universe is baryonic. The remaining
mass is assumed to be a mixture of different forms of yet unknown dark matter 
and dark energy. However, we do have evidence, mainly from the study of large 
scale structures, about the properties of dark matter. 

Cold dark matter (CDM) by definition has a non-relativistic equation of state 
at the beginning of structure formation around the matter-radiation 
equality \cite{Blu}. For successful structure formation an important fraction 
of the dark mass has to be cold dark matter. Although purely baryonic 
matter and hot dark matter (relativistic equation of state at 
matter-radiation equality) models have been ruled out long ago \cite{old}, 
a model with a cosmological constant and baryonic matter only provides a 
good fit to the recent CMB observations \cite{McGaugh}. When 
combined with other cosmological observations it turns out that the small 
sound speed (at photon decoupling) of the baryonic matter 
can only be compatible with the observed multipole moments if the universe 
is closed \cite{BLDM}. Moreover, this model does not provide enough 
power at small scales to explain the observed distribution of galaxies 
\cite{BLDM}. 

The most important feature of CDM is hierarchical structure formation, 
i.~e., small structures form first and grow to larger structures later.
A natural question obviously is: What is the smallest mass scale \cite{ss}
of CDM structures? 

Since the nature of CDM is unknown, the answer to this question will not be 
the same for different CDM candidates. Natural candidates are particles 
that are predicted by extensions of the standard model of particle physics. 
A CDM candidate should be (meta-)stable, (electrical and color) neutral, 
and heavy.The minimal supersymmetric standard model
(MSSM) with the assumption of conserved R-parity provides an excellent 
candidate: the lightest neutralino $\tilde{\chi}_1^{0}$, which probably is 
the lightest supersymmetric particle (LSP) \cite{Jun}.  

The time of kinetic decoupling of CDM depends on the nature of CDM 
\cite{SSW,SH,BFS}. During kinetic decoupling collisional damping is the 
dominant mechanism. Once CDM is fully decoupled from the radiation fluid, 
damping due to free streaming happens. Interesting general considerations 
on damping mechanisms for CDM have been published by Boehm, Fayet, and 
Schaeffer \cite{BFS} recently. 

After the neutralinos decouple chemically (at about $T_{\rm cd} \sim 
M_{\tilde{\chi}}/20$) they remain in kinetic equilibrium due to frequent 
scattering with particles from the radiation fluid. After the QCD transition
(at $\sim 160$ MeV) neutralino-lepton scattering is the most important process.
The neutralinos decouple kinetically once the relaxation time $\tau$ becomes 
comparable with the Hubble time $t_{\rm H} \equiv H^{-1}$, which happens,
depending on the parameters of the MSSM between $10$ MeV and a few $100$ MeV. 
Once collisions of neutralinos with particles from the radiation fluid 
cease, the equation of state becomes nonrelativistic ($P\approx 0$) and 
neutralino matter starts its life as cold dark matter \cite{SSW,SH,CKZ}. 

In the present work we calculate the temperatures of kinetic decoupling 
and last scattering of neutralino CDM for the case of a bino-like neutralino.
A first estimate of 
the kinetic decoupling temperature, based on dimensional arguments, was given 
by Schmid, Schwarz, and Widerin \cite{SSW}, which has been confirmed recently 
by more detailed calculations by Chen, Kamionkowski, and Zhang \cite{CKZ}. 
However, the authors of Ref.~\cite{CKZ} ignored the fact that the relevant 
time scale that has to be compared to the Hubble time is the relaxation time,
rather than the collision time. It was shown explicitly that 
photon-neutralino scattering is suppressed by several orders of magnitudes,
compared to lepton-neutralino scattering.

During the process of kinetic decoupling, collisional damping can smear out 
primordial fluctuations in neutralino CDM below some mass scale $M_{\rm d}$. 
Free streaming gives rise to additional damping below $M_{\rm fs}(t)$, which
depends, in contrast to $M_{\rm d}$, on time. Both damping mechanisms together
give rise to a sharp cut-off in the primordial power spectrum of neutralino 
CDM, that typically lies at $M \sim  10^{-7} M_\odot$ at the time of
matter-radiation equality. We have presented preliminary estimates in 
Ref.~\cite{SH}. In Ref.~\cite{CKZ} it was pointed out that
the estimate of induced damping found in \cite{BFS}, is wrong by several 
orders of magnitude mainly because the cross section for elastic 
scatterings of photons with neutralinos has been overestimated. 

We also show that bulk viscosity, besides shear viscosity, can not be 
neglected (as has been done in \cite{BFS}) in the situation when a 
nonrelativistic component decouples from a radiation fluid. At first sight 
this is a surprising result since bulk viscosity usually goes along with the 
transfer of energy to internal degrees of freedom or with particle production.
None of these mechanisms is available here. However, CDM and radiation have 
to be treated as two separate fluids and the bulk viscosity of the CDM fluid 
just reflects the energy dissipation from the CDM fluid to the radiation 
fluid,
which is however a negligible effect for the radiation fluid since the 
energy density of the CDM fluid is tiny compared to the energy density of 
the radiation fluid at kinetic decoupling. On the other hand the heat 
conduction (which has been considered in \cite{BFS}) can be neglected for the 
CDM fluid. The reason is simple, the neutralinos are too slow.

The paper is organized as follows: A short summary of mass limits and our 
assumptions about the lightest neutralino is given in Sec. II. Then
we review the simplest calculation of chemical decoupling for pedagogical 
reasons and compare that with our detailed calculation of the kinetic 
decoupling and last scattering temperatures (Sec. III). In the following 
section (IV) we introduce CDM as an imperfect fluid, along the lines described
in \cite{Eck,Land,Wei}. The kinetic theory for the description of CDM is 
explained in Sec. V, and the coefficients of transport are calculated in 
Sec. VI. For this purpose we generalize the program by Weinberg \cite{Wei} 
and Straumann \cite{Strau} to the situation of a nonrelativistic component 
that decouples from a relativistic fluid [Weinberg and Straumann treat the 
problem of decoupling of a relativistic component (photons) from a 
nonrelativistic fluid (baryons)]. This finally allows us to calculate the 
damping scale from kinetic decoupling (Sec. VII) and free streaming 
(Sec. VIII). We conclude with a short discussion of the implications of 
our findings. The relevant cross sections are calculated in App. A and some 
useful thermodynamical relations can be found in App. B. 

\section{Neutralinos}

A direct lower limit for the neutralino mass $M_{\tilde{\chi}}$ is provided by 
the LEP experiments, $M_{\tilde{\chi}} > 37$ GeV for any $\tan \beta$ 
and sfermion mass \cite{aleph}. Reasonable assumptions (universal soft 
supersymmetry-breaking scalar masses at some higher scale) within the 
MSSM (constrained MSSM) and taking results of the 
Higgs searches into account raises the lower limit to about $50$ GeV 
\cite{aleph,Ell}. Incorporating also constraints from
$b\rightarrow s\gamma$ decays and assuming that neutralino dark matter is 
cosmologically interesting ($0.1 < \omega_{\tilde{\chi}} < 0.3$) a lower 
limit as high as $M_{\tilde{\chi}} \ge 140 $GeV can be derived \cite{Ell}.
The cosmological upper limit also gives rise to an upper limit on the 
neutralino mass. It is essential to include the effects of the 
next-to-lightest supersymmetric particles and coannihilations, as well as 
the contribution from poles and thresholds properly 
\cite{Ell,GKT,RelicAbundance}. 
A detailed analysis gives $M_{\tilde{\chi}} < 600$ GeV \cite{Ell}.
Since many untested, although reasonable, assumptions go into these limits,
we decided for this work to assume $M_{\tilde{\chi}} > 50$ GeV.  

The neutralinos are linear combinations of the neutral gauginos and 
the two higgsinos of the theory, i.~e., 
\begin{equation}
\tilde{\chi}_1^{0}
=
Z_{11} \tilde{B}^{0} 
+ Z_{12} \tilde{W}_3^{0}
+ Z_{13} \tilde{H}_1^{0}
+ Z_{14} \tilde{H}_2^{0}
\end{equation}
expressed in terms of mass eigenstates. The $Z_{1j}, j\in\{1,2,3,4\}$
are elements of a real orthogonal matrix which diagonalizes the neutralino 
mass matrix. In most of the parameter space of the constrained
MSSM the LSP is a $\tilde{B}^0$. We assume $\tilde{\chi}_1^{0} \equiv 
\tilde{\chi} \approx \tilde{B}^{0}$.

For a pure bino the interaction with a standard model fermion
$F$ is given via the exchange of the related left- or
right-handed sfermion $\tilde{F}_{L,R}$ as follows: 
\begin{equation}
\label{Lint}
{\cal L}_{F \tilde{F} \tilde{\chi}}
=
-\sqrt{2} g \:
\bar{F} \left\{
b_F \; \tilde{F}_L {\cal P}_R 
- 
c_F \; \tilde{F}_R {\cal P}_L
\right\}
+ \rm{h.c.}
\; ,
\end{equation}
where $g$ is the electroweak coupling constant,
${\cal P}_{L,R}$ denotes the left and right
chiral projection operator. The left and right
chiral vertices are given by
\begin{eqnarray}
b_F
&=&
Z_{11} \frac{Y_F}{2} \rm{tan} \theta_W + Z_{12} T_{3F}
\: ,\\
c_F
&=&
Z_{11} Q_F \rm{tan} \theta_W 
\: .
\end{eqnarray}
Here $Y_F$, $T_{3F}$ and $Q_F$
are the weak hypercharge, isospin and electrical charge
of the involved fermions.

\section{Chemical and kinetic decoupling}

There is a large difference between the temperature
of chemical decoupling, $T_{\rm cd}$,  and the temperature of kinetic 
decoupling, $T_{\rm kd}$, of neutralino cold dark matter.
This is a characteristic feature of weakly interacting massive particles 
(WIMPs). Chemical decoupling (freeze-out) fixes the relic abundance of 
neutralinos and therefore the present value of $\Omega_{\rm CDM} h^2$.
Before kinetic decoupling the neutralinos are tightly coupled to radiation, 
after kinetic decoupling the neutralinos acquire the properties of CDM. 
Namely, the neutralinos interact with radiation only via gravity and their 
pressure is negligible compared to their energy density well before 
matter-radiation equality.

Let us first review the process of chemical decoupling, which is a useful 
warm up for the kinetic decoupling that is explained subsequently. We assume
that the neutralino is the bino, which reduces the number of free parameters
to the bino mass, $M_{\tilde \chi}$ and to the universal 
sfermion mass, $M_{\tilde{F}}$. For a more complete picture including
higgsino admixture, thresholds, poles, and coannihilations see 
\cite{Griest,GKT,RelicAbundance}. 

At $T \gg T_{\rm cd}$ neutralinos are kept in chemical equilibrium
with all standard model fermions $F$ in the heat
bath at temperature $T$ via annihilation processes 
$\tilde{\chi} + \tilde{\chi} \leftrightarrow F + \bar{F}$.
{}From (\ref{Lint}) one can calculate the annihilation rate
for $\tilde{\chi} + \tilde{\chi} \rightarrow F + \bar{F}$ 
\cite{Griest},
\begin{eqnarray}
\label{gan}
\Gamma_{{\rm ann}} (T)
&=&
\sum_F \left\langle
v \sigma_{\rm ann} 
\right\rangle (T) \; n_{\tilde{\chi}}(T)
\\  
&=&
\frac{2}{\pi} \; \sum_F
\left(\frac{G_{\rm F} M_W^{\; \; 2}}{M_{\tilde{F}}^{\; \; 2} 
+ M_{\tilde{\chi}}^{\; \; 2}}\right)^2
\Bigg[
(b_F^{\;\;  2}+c_F^{\;\;  2})^2 m_F^2 +
\nonumber \\ 
&+& 4(b_F^{\;\;  4}+c_F^{\;\;  4}) 
\frac{M_{\tilde{F}}^{\; \; 4} + M_{\tilde{\chi}}^{\; \; 4}}
{(M_{\tilde{F}}^{\; \; 2} 
+ M_{\tilde{\chi}}^{\; \; 2})^2} M_{\tilde{\chi}} T
\Bigg]\; n_{\tilde{\chi}}(T) \; .
\label{gan1}
\end{eqnarray}
Here, $\langle \dots \rangle$ denotes thermal averaging
and $v$ is the Moeller-velocity.
In order to obtain (\ref{gan1}) we expanded
$v \sigma_{\rm{ann}}$ for small $m_F/M_{\tilde{\chi}}$ 
and small $v$. More details can be found in appendix A.
Note that the first term in the square brackets 
contributes practically only for top quarks ($F=t$).
However, we will assume below that  $M_{\tilde{\chi}} < m_t$ 
such that the second term will be dominant in our estimate. 
We neglect annihilation of neutralinos into final states
containing gauge and Higgs bosons, such as $\tilde{\chi} \tilde{\chi} 
\leftrightarrow \left\{W W,Z Z,H H, H W, H Z\right\}$,
since these channels are particularly important for Higgsino-like
and mixed-state neutralinos, but are subdominant when
compared to the fermion-antifermion channels in the case
that the neutralino is mostly a gaugino \cite{GKT}.
Since we restrict our attention to a pure bino, there is no contribution from 
diagrams with $Z^0$ exchange at tree level and therefore the $Z^0$ pole does 
not invalidate our estimate below. 

\begin{figure}
\resizebox{0.95 \linewidth}{!}{\includegraphics{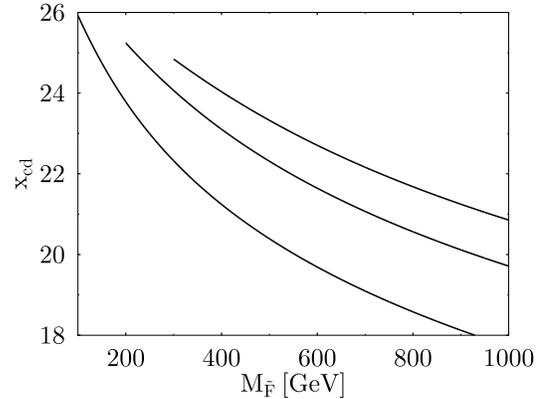}}
\caption{The chemical decoupling as a function of the sfermion 
mass $M_{\tilde{F}}$ for three values of the neutralino mass 
$M_{\tilde{\chi}} = 50, 100, 150$ GeV (increasing from bottom to top).
\label{fig1}}
\end{figure}

As the universe expands the temperature eventually falls below the 
neutralino mass $M_{\tilde{\chi}}$ and the number density $n_{\tilde{\chi}}$
of neutralinos decreases exponentially. Once the annihilation rate
$\Gamma_{\rm ann}$ becomes comparable to the expansion rate $H$ of the 
universe neutralinos no longer find other neutralinos to annihilate. 
We use the condition $\Gamma_{\rm ann} = H$ to define the temperature of 
chemical decoupling $T_{\rm cd}$. Solving this equation iteratively yields 
($x\equiv M_{\tilde{\chi}}/T$) 
\begin{eqnarray}
\label{xf}
x_{\rm cd}^{(0)}
&=&
{\rm ln}
\left[
1.6 \times 10^{-4} 
\frac{
M_{\rm Pl} (M_{\tilde{F}}^4 + M_{\tilde{\chi}}^4) M^3_{\tilde{\chi}}}
{(M_{\tilde{F}}^2+M_{\tilde{\chi}}^2)^4}
\right]
\; ,
\nonumber \\
x_{\rm cd}^{(1)}
&\approx&
x_{\rm cd}^{(0)} - \frac{1}{2} \;  {\rm ln} \; x_{\rm cd}^{(0)}
\; ,
\end{eqnarray} 
as long as the bino mass is well below the top mass, but large compared to the 
bottom mass. In deriving (\ref{xf}) we assumed equal masses for all 
sfermions. Exploring the parameter space of the MSSM 
we typically find $x_{\rm cd} \approx 25$, cf. figure \ref{fig1}. 

%%%%%%%%%%%%%%%%%%%%%%%%%%%%%%%%%%%%%%%%%%%%%%%%%%%%%%%%%%%%%%
%%%%%%%%%%%%%%%%%%%%%%%%%%%%%%%%%%%%%%%%%%%%%%%%%%%%%%%%%%%%%%%

\begin{figure}
\resizebox{0.95 \linewidth}{!}{\includegraphics{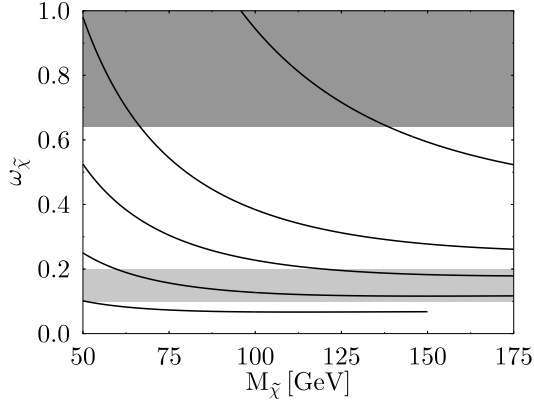}}
\caption{The relic abundance of neutralinos expressed by $\omega_{\tilde \chi} 
= \Omega_{\tilde \chi} h^2$ as a function of the neutralino mass 
$M_{\tilde{\chi}}$ for 
different values of the sfermion mass $M_{\tilde{F}} = 150, 200, 250, 
300, 400$ GeV. The sfermion mass
increases from the bottom to the top. The dark shaded region is excluded 
by the conservative assumptions: $\Omega \leq 1$ and $h < 0.8$. The light 
shaded region indicates typical values of $\omega$ in a $\Lambda$CDM model.
\label{fig2}}
\end{figure}

The relic abundance of neutralinos is now easily obtained as 
$n_{\tilde{\chi}}(T_0) = n_{\tilde{\chi}}(T_{\rm cd}) s(T_0)/s(T_{\rm cd})$, 
where $T_0 = 2.725$ K and $s$ denotes the entropy density of the universe. 
It is a good approximation to use the equilibrium distribution for the number 
density at $T_{\rm cd}$, although in a more advanced treatment the 
corresponding kinetic equation should be solved. From the number density 
$n_{\tilde{\chi}}(T_0)$ we may easily compute $\omega_{\tilde{\chi}} 
\equiv \Omega_{\tilde{\chi}} h^2$, which is plotted in Fig.~\ref{fig2} as 
a function of $M_{\tilde{\chi}}$ for typical values of the sfermion mass.

Below $T_{\rm cd}$ the neutralinos are kept in local thermal equilibrium
via elastic scattering processes $\tilde{\chi} + F \rightarrow 
\tilde{\chi} + F$. After the QCD phase transition only leptons $L$ remain
as scattering partners for the neutralinos. We neglect scatterings 
with pions, which is important for $T > m_\pi$ only. It will turn out that in
most cases $T_{\rm kd} \ll m_\pi$. Scattering with nucleons is not important 
due to the tiny number density of baryons. {}From (\ref{Lint}) one can 
calculate the rate of elastic scatterings 
$\tilde{\chi} + L \rightarrow \tilde{\chi} + L$ \cite{Griest}. We find
\begin{eqnarray}
\label{gel}
\Gamma_{\rm el} 
&=&
\sum_L 
\left\langle v \sigma_{\rm el} (E_L) \right\rangle (T) \; n_L(T)
\\ 
&=&
\frac{288}{\pi} \; \sum_L (b_L^{\; \; 4} + c_L^{\; \; 4})
\left(
\frac{G_{\rm F} M_W^{\; \; 2}}{M_{\tilde{L}}^{\; \; 2} 
- M_{\tilde{\chi}}^{\; \; 2}}
\right)^2 T^2 \; n_L (T) \; .
\label{gel1}
\end{eqnarray}
$E_L$ denotes the energy and $n_L$ the number density of the leptons.
In deriving (\ref{gel1}) we approximate the Mandelstam variable 
$s\approx M_{\tilde{\chi}}^{\; \; 2} + 2 M_{\tilde{\chi}} E_L$.
Note that the Moeller velocity in this case is $v\approx 1$ to a very good 
approximation.

%%%%%%%%%%%%%%%%%%%%%%%%%%%%%%%%%%%%%%%%%%%%%%%%%%%%%%%%%%%%%%%
%%%%%%%%%%%%%%%%%%%%%%%%%%%%%%%%%%%%%%%%%%%%%%%%%%%%%%%%%%%%%%

In analogy to the chemical decoupling the condition $\Gamma_{\rm el} = H$ 
defines the temperature at the time the last elastic interactions between 
neutralinos and the rest of the Universe take place. This last scattering 
temperature is given by
\begin{equation}
T_{\rm ls} = \left[8.7 \times 10^{-3} 
{m_{\rm Pl} \over (M_{\tilde{L}}^2 - M_{\tilde{\chi}}^2)^2}\right]^{-1/3}.
\end{equation}
Typical values are $1$ MeV to $10$ MeV, e.g., $T_{\rm ls} = 2.3 (2.5)$ MeV for
$M_{\tilde{\chi}} = 100 (150)$ GeV and $M_{\tilde{L}} = 200 (250)$ GeV.
However, this is {\emph not} the temperature
at which neutralinos decouple kinetically. The
kinetic decoupling temperature is defined through
the relaxation time $\tau$, rather than by the 
collision time $\tau_{\rm coll} = 1/\Gamma_{\rm el}$. 
This can be easily understood by the following argument.  

The relaxation time $\tau$, i.e., the time neutralinos need to 
return to local thermal equilibrium after a deviation from it,
can be estimated from the typical number of scatterings that is needed to 
change the momentum of the neutralino significantly. The typical 
momentum transfer in a single elastic scattering event is tiny compared 
to the average momentum of the neutralinos. This is easily seen from 
the averaged Mandelstam variable $t$, 
\begin{equation}
\left(\Delta p_{\tilde{\chi}}\right)^2 
\equiv
-\frac{1}{\sigma_{\rm el}} \; 
\int {{\rm d}\sigma_{\rm el}\over {\rm d} t} \; t {\rm d} t
=
2 \; E_L^2.
\end{equation}
The leptons are kept in local thermal equilibrium
through the frequent interactions among themselves and
the equipartition theorem gives $E_L = 3/2 \; T$.
Comparing the rms momentum transfer with the typical
neutralino momentum $p_{\tilde{\chi}}$ we find
$\Delta p_{\tilde{\chi}}/p_{\tilde{\chi}}=
\sqrt{3/2}\; T/M_{\tilde{\chi}}\ll 1$.
This means that a huge number $N(T)$ of elastic scatterings 
is needed to keep or to establish thermal equilibrium,
$N(T) = p_{\tilde{\chi}} / \Delta p_{\tilde{\chi}}
= \sqrt{3/2} \; M_{\tilde{\chi}}/T$. 
We can now estimate the relaxation time as
\begin{equation}
\tau (T)
\approx 
{\sqrt{\frac{2}{3}}}\; \frac{M_{\tilde{\chi}}}{T} 
\tau_{\rm coll}\; .
\end{equation}
Note that $\tau (T)\sim 1/T^6$. 

\begin{figure}
\resizebox{0.95 \linewidth}{!}
{\includegraphics{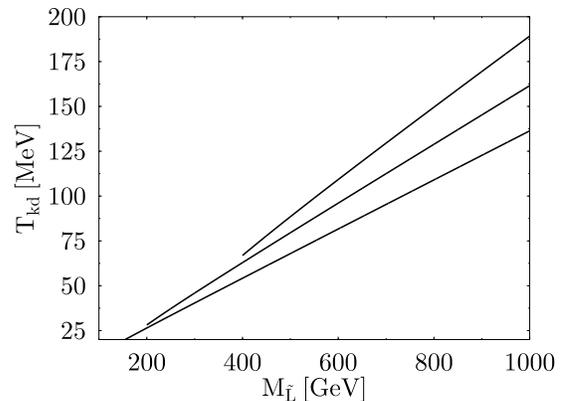}}
\caption{The temperature of kinetic decoupling of neutralinos from
radiation as a function of the sfermion mass for 
$M_{\tilde{\chi}} = 50, 100, 200$ GeV (bottom to top).
\label{fig3}}
\end{figure}

The kinetic decoupling of the neutralinos happens when the relaxation time 
$\tau$ becomes comparable to the Hubble time $1/H$. We denote the 
corresponding temperature by $T_{\rm kd}$, which is given by 
\begin{equation}
T_{\rm kd} = \left[1.2 \times 10^{-2}
{m_{\rm Pl} \over M_{\tilde{\chi}} 
(M_{\tilde{L}}^2 - M_{\tilde{\chi}}^2)^2}\right]^{-1/4},
\end{equation}
where we assumed that all leptons except the tau are relativistic,
but we neglected the contribution of pions which are important at temperatures
of about $130$ MeV. Above the QCD phase transition at about $160$ MeV much
more interaction partners are available and our formula should be modified. 
Exploring the parameter space of the MSSM we typically find that 
$T_{\rm kd}$ is of the order $10$ MeV to $100$ MeV, cf. figure \ref{fig3}.
For $M_{\tilde{\chi}} = 100 (150)$ GeV and $M_{\tilde{L}} = 200 (250)$ GeV we 
find $T_{\rm kd} = 28 (36)$ MeV, whereas the chemical decoupling for the same
set of parameters happens at $T_{\rm cd} = 4.0 (5.9)$ GeV, a difference of 
more than two orders of magnitude.
%%%%%%%%%%%%%%%%%%%%%%%%%%%%%%%%%%%%%%%%%%%%%%%%%%%%%%%%%%%%%%%%%%%%%%%%%
%%%%%%%%%%%%%%%%%%%%%%%%%%%%%%%%%%%%%%%%%%%%%%%%%%%%%%%%%%%%%%

The large difference between $T_{\rm cd}$ and $T_{\rm kd}$ is mainly due 
to the different target densities in the annihilation [Eq.~(\ref{gan})] and 
elastic scattering rates [Eq.~(\ref{gel})]. For annihilations the target 
density is given by the number density $n_{\tilde{\chi}}$ of neutralinos 
in the universe. The number density of neutralinos is suppressed by the 
Boltzmann factor at chemical decoupling. In contrast the target density for 
elastic scattering processes is given by the number density of all 
relativistic leptons.

During the cooling from $T_{\rm cd}$ to $T_{\rm kd}$
the leptons behave as a perfect radiation fluid
which tries to keep the neutralinos in thermal equilibrium through
elastic scattering processes. The neutralinos on the other hand may be 
described as a nonrelativistic, imperfect fluid.

\section{CDM as a fluid}

For temperatures $T > T_{\rm cd}$ the particle content
of the universe may be described by a single radiation fluid
which is in local thermal equilibrium.  
For temperatures $T_{\rm kd}<T<T_{\rm cd}$ the radiation fluid
is tightly coupled to the CDM fluid. Hence, both fluids have the
same temperature and velocity four-vectors. 
Around $T_{\rm kd}$ the CDM fluid starts to decouple kinetically from the 
radiation fluid and becomes an imperfect fluid. The departure
from local thermal equilibrium is generated by dissipation,
i.e. by shear and bulk viscosity (we show below that the coefficient of
heat conduction vanishes). 
For temperatures $T<T_{\rm kd}$ both fluids are decoupled
and the CDM fluid is freely streaming. Since
$\Omega_{\rm{CDM}} = (a/a_{\rm{eq}}) \; \Omega_{\rm{rad}} 
\ll \Omega_{\rm{rad}}$ for $T\gg T_{\rm{eq}}$,
the radiation fluid remains in local thermal equilibrium throughout the 
decoupling process.

The current density and the energy-momentum tensor of the radiation fluid (R)
are given by 
\begin{eqnarray}
J_{\rm{R}}^{\; \mu}
&=&
n_{\rm{R}} V^\mu \; , \\
T_{\rm{R}}^{\; \mu\nu}
&=&
\rho_{\rm{R}} \; V^\mu V^\nu - P_{\rm{R}} \; h^{\mu\nu}
\; .
\end{eqnarray}
Here, $n_{\rm{R}}$, $\rho_{\rm{R}}$, and $P_{\rm{R}}$ are the number 
density, the energy density, and the pressure of the radiation fluid 
respectively. $V$ is the velocity four-vector with $V^2=1$. 
$h^{\mu\nu} = g^{\mu\nu} - V^\mu V^\nu$ is the projection operator 
on the plane perpendicular to $V$. The radiation fluid variables only 
depend on the temperature of radiation, $T_{\rm R}$, since
there are no 
relevant conserved quantum numbers besides R-parity, which is taken into 
account in the CDM fluid.

The current density and the energy-momentum tensor
of the imperfect CDM fluid can be written as \cite{Eck,Land,Wei}
\begin{eqnarray}
\label{lt1}
J^\mu &=& 
n U^\mu + J^{(1)\;\mu} \ ,\\
\label{lt2}
T^{\mu\nu} &=&
\rho(T,n) U^\mu U^\nu - P(T,n) h^{\mu\nu} + T^{(1)\;\mu\nu} \ .
\end{eqnarray}
$n$, $U$, $\rho(T,n)$, and $P(T,n)$ are the number density, four-velocity,
energy density, and pressure of the CDM fluid, respectively.  We omit 
the subscript $\tilde{\chi}$ for the CDM component in Secs. IV, V, and VI, 
since the results of these sections hold true for more general forms of WIMP 
CDM. The projection $h$ is orthogonal to $U$ here. We do not introduce two 
different symbols in the following, because it is always clear from the 
context to which velocity $h$ is referring. For the CDM
fluid $T$ and $n$ are independent variables, since R-parity, i.e., the number
of neutralinos, is conserved. $T$ is not necessarily identical to $T_{\rm R}$,
although this is the case when both fluids are in thermal equilibrium. 
In the adiabatic limit all space-time gradients are negligible, i.e.,
$J^{(1)} = 0, \; T^{(1)} = 0$, and the CDM fluid has the same
temperature and the same velocity as the radiation fluid, $U=V$. 

For an imperfect fluid described by (\ref{lt1}) and (\ref{lt2})
number density, energy density and velocity  are not defined uniquely. 
To fix this ambiguity we define the number and energy density by 
\begin{eqnarray}
n
\label{c1}
&\equiv&
U_\mu J^\mu
\; , \\
\rho (T,n)
\label{c2}
&\equiv&
U_\mu U_\nu T^{\mu\nu}
\; ,
\end{eqnarray}
such that the hydrodynamic and thermodynamic definitions of $n$ and $\rho$
coincide. The velocity is fixed to be the comoving velocity of the CDM 
particles 
\begin{equation}
U^\mu
\label{c3}
\equiv
\left( J_\lambda J^\lambda \right)^{-\frac{1}{2}} \; J^\mu
\; .
\end{equation}
This choice of the velocity corresponds to the one of Eckart \cite{Eck}, 
and was applied to relativistic fluids by Weinberg \cite{Wei}. An 
alternative would be the choice of Landau \cite{Land}, where the velocity 
is fixed to coincide with the velocity of the energy-momentum flow.
In other words (\ref{c1}) requires $J^{(1)}$ to be perpendicular to $U$.
In the same sense (\ref{c2}) requires $T^{(1)}$ to project on the plane
perpendicular to $U$.  Condition (\ref{c3}) means that $J^{(1)}$ has to 
vanish.
With this choice
\begin{eqnarray}
J^\mu
&=&
n U^\mu
\; ,\\
T^{\mu\nu}
&=&
\rho(T,n) U^\mu U^\nu - P(T,n) h^{\mu\nu}
+ T^{(1)\; \mu\nu}
\;. 
\end{eqnarray}
The construction of $T^{(1)}$ from a first-order formalism is given in 
\cite{Wei}. The starting point is that $T^{(1)}$ has to be expressed in 
the equilibrium variables and their gradients. The basic observation is that 
the variation of the entropy per particle, $\sigma$, along the adiabatic flow
is generated by the change of $T^{(1)}$ in the same
space-time direction, i.e.,
\begin{equation}
\label{ent}
nT \; \dot{\sigma}
=
- U_\mu T^{(1)\; \mu\nu}_{\; \; \; \; \; \; \; \; \; , \nu}
\; ,
\end{equation}
where the dot denotes the hydrodynamic derivative, $\dot{()} \equiv 
U^\mu ( )_{, \mu}$. Equivalently we may write for the entropy current 
four-vector 
$S^\mu \equiv n \sigma U^\mu + 1/T \; U_\lambda T^{(1)\; \lambda\mu}$
(as can be easily seen in the comoving frame, $S^0 = n \sigma = s$ is indeed 
the entropy density and $T S^i = T^{(1)\; 0i}$ is the non-adiabatic 
contribution to the energy-momentum flow, which is the heat flow)  
\begin{equation}
\label{Ent}
T^2\; S^\lambda_{\; ,\lambda}
=
- \left( U_\mu T_{\; ,\nu} 
- T U_{\mu \; ,\nu} \right) \; T^{(1)\; \mu\nu}
\; .
\end{equation}
As a consequence only space-time derivatives of $T$
and $U$ can occur in $T^{(1)}$ in order to keep
the rate of entropy production positive for
all fluid configurations.

The perturbed energy-momentum tensor
may be expressed in terms of the heat-flow 
vector 
\begin{equation}
Q_\mu
=
T_{\; ,\mu} - T\dot{U}_\mu
\end{equation}
and the traceless shear tensor
\begin{equation}
W_{\mu\nu}
=
U_{\mu\; ,\nu} + U_{\nu\; ,\mu}
-\frac{2}{3} \; g_{\mu\nu} \;  U^\lambda_{\; ,\lambda}
\; .
\end{equation}
With these abbreviations we write \cite{Wei} 
\begin{eqnarray}
T^{(1) \; \mu\nu}
&=& \zeta \; h^{\mu\nu} \; U^{\lambda}_{,\lambda} +
\eta \; h^{\mu\rho} h^{\nu\sigma} \; W_{\rho\sigma} -
\nonumber \\
\label{emt1}
& & \chi \; (h^{\lambda\mu} U^{\nu} + h^{\lambda \nu} U^{\mu}) \;
Q_\lambda
\; ,
\end{eqnarray}
where $\zeta$, $\eta$ and $\chi$ are the coefficients
for bulk viscosity, shear viscosity and
heat conduction.

These parameters need to be calculated in the framework
of a non-equilibrium theory. 

\section{Kinetic description of CDM}

In this and the following section we generalize a method by Straumann 
\cite{Strau} to calculate the coefficients of transport for a species of
massive particles that decouple kinetically. In Ref.~\cite{Strau} the problem
of the decoupling of radiation quanta was treated.

Let $F(p,x)$ be the distribution function of neutralinos.
$F(p,x)$ is normalized in such a way that
$F(p,x) \; {\rm d}^3p {\rm d}^3 x$ gives the number of quanta in the
volume ${\rm d}^3 x$ centered at the space-time point $x$
and three-momentum within ${\rm d}^3 p$.
We assume that the neutralinos are close to thermal equilibrium
and make the ansatz 
\begin{equation}
\label{ans}
F=F^{(0)}+F^{(1)} 
\hspace{0.5cm} \mbox{with} \hspace{0.5cm} 
\left| F^{(1)}\right| \ll F^{(0)}
\; ,
\end{equation}
where
\begin{equation}
F^{(0)}(p,x)
=
\frac{g}{(2\pi)^3} \; 
\frac{1}{\exp\left(\frac{p\cdot V}
{T_{\rm R}}-\alpha\right) \pm 1}
\; .
\end{equation}   
Here, $T_{\rm{R}}$ is the local temperature of the radiation fluid 
and $\alpha$ is the local pseudo chemical potential of the neutralino.
To first order in the collision time, $F^{(1)}$ is a solution
of the kinetic equation 
\begin{equation}
\label{keq}
(p\cdot \partial) \;  F^{(0)}
=
L[F^{(1)}]
\; .
\end{equation}
$L$ is supposed to be a linear functional in $F^{(1)}$.
In the linear regime one often uses 
$L[F^{(1)}] = -\omega \tau^{-1} \; F^{(1)}$
as a realistic model for the collision integral.

In consideration of Eckarts approach to the hydrodynamics of
imperfect fluids \cite{Eck} we introduce a four-vector
perpendicular to $V$,
\begin{equation}
\label{ndef}
n^{\mu}
=
|\vec{p}|^{-1} \left(p^{\mu} - \omega V^{\mu}\right)
\end{equation}
with $\omega=p\cdot V$ such that $F^{(1)}$ can be considered as
a function of $\omega$, $n$ and $x$ or equivalent
as a function of the projection of $p$ in the direction
of $V$ and perpendicular to it.

Following \cite{Mas} we may now expand $F_P^{(1)}(\omega,n,x)$ into
polynomials in $n$
\begin{eqnarray}
F^{(1)}(\omega,n,x)
&=& A(\omega,x)+B_{\mu}(\omega,x)n^\mu+ \nonumber\\
& & C_{\mu\nu}(\omega,x)
\left(n^{\mu}n^{\nu}+\frac{1}{3} \; h^{\mu\nu} \right)
+ ...
\; .
\label{F1}
\end{eqnarray}
It is clear from the kinetic equation (\ref{keq})
that we need to know how the functional $L$
operates on $F^{(1)}$. In order to solve this problem we note
that $F$ is defined to be invariant under Lorentz-transformations.
Let $G_x$ be the group of all Lorentz-transformations
leaving $V(x)$ invariant at every space-time point $x$,
i.e. $G_x$ is the little group with respect to $V$.
$G_x$ is isomorphic to the Lie-group $SO(3)$.
Since $F$ is invariant under $G_x$
at every space-time point $x$, $F^{(1)}$ is invariant and
(\ref{F1}) is an expansion into irreducible polynomials
with respect to $G_x$. From equation (\ref{keq}) 
it follows that the linear functional $L$
is a scalar with respect to $G_x$. Therefore
it operates on the irreducible subspace spanned 
by the polynomials in (\ref{F1})
as a multiple of the identity. Thus we can write
\begin{eqnarray}
L[F^{(1)}]
&=& -\omega \left[\kappa_0 A + \kappa_1 B_{\mu} n^{\mu} + \right. 
\nonumber \\
& & \left. + \kappa_2 C_{\mu\nu} \left(n^{\mu}n^{\nu}+
\frac{1}{3} \; h^{\mu\nu}\right) \right]
+ ...
\; ,
\label{entw}
\end{eqnarray}
where the $\kappa_j$ $(j\in \{0,1,2\})$ 
are functions of $\omega$ and $x$ only.
Note that in the case of the model for the collision integral
discussed above $\kappa_j = \tau^{-1}$ for all 
$j\in \{1,2,3\}$. 

Next we derive expressions for $A$, $B_{\mu}$ and $C_{\mu\nu}$
in terms of $V$, $T_{\rm{R}}$, $\alpha$ and 
$\kappa_j$ using the kinetic equation 
$(\ref{keq})$.
In order to do this we have to define a measure ${\rm d}\Omega_V$
on the two dimensional surface $S=\{p: p^2=M^2, p^0 > 0$
and $\omega=$const.$\}$. ${\rm d}\Omega_V$ is normalized such that
\begin{equation}
\frac{1}{4\pi} \int_S {\rm d}\Omega_V = 1
\; .
\end{equation}
The irreducible polynomials in (\ref{keq})
are orthogonal with respect to ${\rm d}\Omega_V$
and are normalized as follows
\begin{eqnarray}
&&\frac{1}{4\pi} \int_S {\rm d}\Omega_V n^{\mu} n^{\nu}
=
-\frac{1}{3} \; h^{\mu\nu}
\; ,\\ \nonumber \\
&&\frac{1}{4\pi} \int_S {\rm d}\Omega_V
\left(n^{\mu} n^{\nu} + \frac{1}{3} \; h^{\mu\nu} \right)
\left(n^{\alpha} n^{\beta} + \frac{1}{3} \; h^{\alpha\beta} \right)
= \nonumber \\
&& \qquad 
\frac{1}{15} \; h^{\{\mu\nu} \; h^{\alpha\beta\}}
-\frac{1}{9} \; h^{\mu\nu} \; h^{\alpha\beta} \; .
\end{eqnarray}
Now it is possible to project out every tensor
in the expansion of $L[F^{(1)}]$, equation (\ref{F1}).
Taking moments of (\ref{keq}) and using (\ref{F1})
we obtain
\begin{eqnarray}
\label{A}
A
&=&
\frac{1}{\kappa_0}\frac{\omega\Phi'}{T_{\rm{R}}}
\left[\frac{\dot{T_{\rm{R}}}}{T_{\rm{R}}}   
+ \frac{1}{3}\left(1-\frac{m^2}{\omega^2}\right)^2 V^\lambda_{,\lambda}
+ \frac{T_{\rm{R}}}{\omega}\dot{\alpha}\right]\; ,\\
\label{B}
B_{\mu}
&=&
\frac{1}{\kappa_1}\frac{\omega\Phi'}{T_{\rm{R}}}
\left(1-\frac{m^2}{\omega^2}\right)^{\frac{1}{2}} h_\mu^{\;\lambda}
\left[\frac{1}{T_{\rm{R}}} Q_\lambda + \frac{T_{\rm{R}}}{\omega}
\alpha_{,\lambda}
\right]\; , \\
\label{C}
C_{\mu\nu}
&=&
-\frac{1}{\kappa_2}\frac{\omega \Phi'}{T_{\rm{R}}} 
\frac{1}{2}\left(1-\frac{m^2}{\omega^2}\right)^{2} 
h_\mu^{\;\lambda} h_\nu^{\;\gamma} W_{\lambda\gamma}\; .
\end{eqnarray}
Here, $\Phi'(\omega/T_{\rm{R}}-\alpha)$ denotes
the external derivative of $F^{(0)}(\omega,x)$.
In calculating 
the integrals we replaced $(p\cdot\partial) \; T_{\rm{R}}$
with $\omega (V\cdot\partial) \; T_{\rm{R}}$ and
$p_{\beta} (p\cdot \partial) \; V^{\beta}$
with
$|\vec{p}|^2 n_{\beta} (n\cdot \partial) \; V^{\beta}$. 

The coefficients $(\ref{A})$ and $(\ref{B})$
depend on the variations of $T_{\rm{R}}$ and $\alpha$
along the adiabatic flow and the directional
derivative of $\alpha$ in the plane perpendicular to
the adiabatic flow. In order to make sure that the
rate of entropy production along the adiabatic flow
is positive for all kinematical configurations
these derivatives need to be proportional
to space-time gradients of $T_{\rm{R}}$ and $V$.
Using the adiabatic relations derived in Appendix B
we find
\begin{eqnarray}
\label{A1}
A
&=&
-\frac{1}{\kappa_0}\frac{\omega\Phi'}{T_{\rm{R}}} \times \\
&& 
\left[
\left(\frac{\partial P}{\partial \rho}\right)_n
- \frac{1}{3} \left(1-\frac{m^2}{\omega^2}\right)^2 
+ \omega^{-1} \left(\frac{\partial P}{\partial n}\right)_\rho 
\right] V^\lambda_{\; ,\lambda}, \nonumber \\
\label{B1}
B_{\mu}
&=&
\frac{1}{\kappa_1}\frac{\omega\Phi'}{T_{\rm{R}}}
\left(1 -\frac{m^2}{\omega^2}\right)^{\frac{1}{2}}
\left[
\frac{\omega}{T_{\rm{R}}}-\frac{w}{nT_{\rm{R}}}
\right] 
h_\mu^{\; \lambda} Q_\lambda, \\ 
\label{C1}
C_{\mu\nu}
&=&
-\frac{1}{\kappa_2} \frac{\omega\Phi'}{T_{\rm{R}}} 
\frac{1}{2} \left(1-\frac{m^2}{\omega^2}\right)^{2} 
h_\mu^{\; \lambda} h_\nu^{\; \gamma}
W_{\lambda\gamma},
\end{eqnarray}
with the enthalpy $w = \rho + P$.

\subsection{Current density}

In kinetic theory the current density of neutralinos is given by
\begin{equation}
\label{pcd}
J^{k\;\mu} 
= \int 
\frac{{\rm d}^3p}{p^0} \; p^{\mu} \; F(\omega, n, x) \; .
\end{equation}
Considering our ansatz (\ref{ans}), we may write 
\begin{equation}
\label{anj}
J^{k\;\mu}
=
J^{k (0)\; \mu} + J^{k (1) \; \mu}, 
\end{equation}
with the definitions 
\begin{eqnarray}
J^{k (0)\; \mu} 
&\equiv& 
\int {{\rm d}^3 p\over p_0} p^\mu F^{(0)}(\omega, x) = n^{k} V^\mu
\; , \\
n^{k}
&=&
4\pi \int_M^\infty {\rm d} \omega (\omega^2 -M^2)^{1/2} \omega 
\Phi\; , 
\end{eqnarray}
and
\begin{equation}
\label{kcd}
J^{k (1)\; \mu} 
\equiv
\int {{\rm d}^3 p\over p_0} p^\mu F^{(1)}(\omega, n, x) =
\Delta n V^{\mu} + J_{\rm diff}^\mu
\; .
\end{equation}
$\Delta n$ is generated by the coefficient
$A$ whereas $J_{\rm diff}$ is generated by $B$:
\begin{eqnarray}
\Delta n  &=& 4 \pi \int_M^\infty {\rm d} \omega 
(\omega^2-M^2)^{1/2} \omega \; A
\; ,\\
J_{\rm diff}^\mu & = & - {4 \pi\over 3} 
\int_M^\infty {\rm d} \omega (\omega^2 -M^2)  \; B^\mu
\; .
\end{eqnarray}
Let us rewrite the above expressions with help of the following notation:
\begin{equation}
f^{(i,j)}_a(T_{\rm{R}},\alpha;x) \equiv 
- 4 \pi \int_M^\infty {\rm d} \omega (\omega^2 - M^2)^{i/2}
\omega^j {\Phi'\over T_{\rm{R}} \kappa_a} \ .
\end{equation}
Note that the mass dimension of $f^{(i,j)}_a$ is simply $i+j$.
In terms of these functions we obtain 
\begin{eqnarray}
&&\Delta n
= \nonumber \\ 
&&\left[
f_0^{(1,2)} \left(\frac{\partial P}{\partial\rho}\right)_n
- f_0^{(3, 0)} \frac{1}{3} 
+ f_0^{(1, 1)} \left(\frac{\partial P}{\partial n}\right)_\rho
\right] V^\lambda_{\; , \lambda}, 
\label{dn} 
\\ 
\label{jdi}
&&J_{\rm diff}^\mu 
=  
\frac{1}{3 T_{\rm{R}}}  
\left[f_1^{(3, 0)} - f_1^{(3, -1)}\frac{w}{n}\right]
h^{\mu \lambda} Q_\lambda.
\end{eqnarray}

\subsection{Energy-momentum tensor}

The energy-momentum tensor of neutralino CDM is given by 
\begin{equation}
T^{k\;\mu\nu}
=
\int \frac{{\rm d}^3p}{p_0} \; p^{\mu} p^{\nu} \; F(\omega, n, x)
\; .
\end{equation}
Again, in consideration of our ansatz we may write
\begin{equation}
\label{emtk}
T^{k \;\mu\nu}
=
T^{k (0)\;\mu\nu} + T^{k(1)\;\mu\nu}\; ,
\end{equation}
with the definitions
\begin{eqnarray}
T^{k (0)\;\mu\nu}
&\equiv&
\int \frac{{\rm d}^3p}{p_0} \; p^{\mu} p^{\nu} \; F^{(0)}(\omega, x)
\nonumber \\ 
&=&
\rho^k \; V^\mu V^\nu - P^k\; h^{\mu\nu}
\; ,\\
\rho^k
&=& 4\pi \int_M^\infty {\rm d} \omega (\omega^2 -M^2 )^{1/2} \omega^2 
\Phi \; , \\
P^k
&=& {4\pi\over 3} \int_M^\infty {\rm d} \omega (\omega^2 -M^2 )^{3/2} 
\Phi\; , 
\end{eqnarray}
and
\begin{equation}
T^{k (1)\; \mu\nu}
\equiv
T_A^{(1)\; \mu\nu}+
T_B^{(1)\; \mu\nu}+
T_C^{(1)\; \mu\nu}
+\dots \ .
\end{equation}
The labels $A,B,C$ indicate which tensor in the expansion
(\ref{F1}) gives rise to the extra contribution.
{}From (\ref{F1}) and (\ref{A1}) -- (\ref{C1}) we find 
\begin{eqnarray}
\label{TA}
T_A^{(1)\;\mu\nu} &=&
\Delta \rho \; V^\mu V^\nu - \Delta P \; h^{\mu\nu}
\; ,\\ \nonumber\\
\label{TB}
T_B^{(1)\;\mu\nu} 
&=& 
\frac{1}{3T_{\rm{R}}}
\left[
f_1^{(3,1)} 
-f_1^{(3,0)} \; \frac{w}{n}
\right]\;
V^{\{\mu} h^{\nu\}\lambda} \; Q_\lambda
\; ,\\ \nonumber\\
T_C^{(1)\;\mu\nu} 
\label{TC}
&=& \frac 1 {15} f_2^{(5,-1)}
h^{\mu\lambda} h^{\nu\gamma} \; W_{\lambda\gamma}
\; ,
\end{eqnarray}
with
\begin{eqnarray}
&&\Delta \rho
= \\ \nonumber \\
&&\left[
f_0^{(1,3)} \left(\frac{\partial P}{\partial\rho}\right)_n
-  f_0^{(3,1)} \frac{1}{3} 
+f_0^{(1,2)} \left(\frac{\partial P}{\partial n}\right)_\rho
\right] \;
V^\lambda_{\; ,\lambda}
\; ,\nonumber \\ \nonumber \\ 
&&\Delta P
\label{deP}
= \\ \nonumber \\
&&\frac{1}{3} 
\left
[f_0^{(3,1)} \left(\frac{\partial P}{\partial\rho}\right)_n 
- f_0^{(5,-1)} \frac{1}{3}
+f_0^{(3,0)} \left(\frac{\partial P}{\partial n}\right)_\rho
\right] \; 
V^\lambda_{\; ,\lambda}
\; . \nonumber
\end{eqnarray}

\section{Coefficients of transport}

In the following we calculate the coefficients of bulk and shear viscosity
and the coefficient of heat conduction for neutralino CDM starting from the
kinetic description. 

In Sec. III we have introduced Eckart's approach to describe imperfect fluids
\cite{Eck}. 
The number density and the energy density of the CDM fluid coincide with the 
corresponding quantities in the adiabatic limit, see Eqs. (\ref{c1}) and 
(\ref{c2}), and the velocity of the CDM fluid is fixed via the particle 
current, see Eq. (\ref{c3}). These definitions together with the required 
space-time symmetries and the second law of thermodynamics determine the 
most general structure of $J^{(1)}$ and $T^{(1)}$, see \cite{Wei}, as given
in Eq.~(\ref{emt1}).

To compare the kinetic description from the previous section 
with the approach of Eckart it is necessary that the conditions 
(\ref{c1})-(\ref{c3}) are fulfilled. Instead we find for the kinetic 
description
\begin{eqnarray}
V_\mu J^{k\; \mu}
\label{p1}
&=& n^k + \Delta n 
\; ,\\
V_\mu V_\nu T^{k \; \mu\nu}
\label{p2}
&=&
\rho^k + \Delta \rho
\, ,\\
h^{\mu\lambda} J_\lambda
\label{p3}
&=&
J_{\rm{diff}}^\mu
\; .
\end{eqnarray}
Due to the non-equilibrium dynamics, the kinetic number density and energy 
density do not coincide with Eckart's definitions, (\ref{p1}) and (\ref{p2}) 
are in conflict with (\ref{c1}) and (\ref{c2}). Equation (\ref{p3}) shows 
the existence of a diffusion current in the plane perpendicular to $V$. As a 
consequence the current density four-vector does not point to the space-time 
direction that is required by the approach of Eckart (\ref{c3}). In the 
following we consider the temperature and the number density to be the 
independent thermodynamical variables.

Let us first establish the link between the current in the kinetic and the 
hydrodynamic descriptions. As a first step we make a transformation of the
velocity, such that the diffusion current vanishes,
\begin{equation}
\label{trafo}
V\rightarrow
U - (n^k + \Delta n)^{-1} J_{\rm{diff}}
\; ,
\end{equation}
which allows us to write
\begin{equation}
J^{\mu} = (n^k + \Delta n) U^{\mu} = n U^\mu\; ,
\end{equation}
from comparison with (\ref{c1}). 

Let us now turn to the energy-momentum tensor. The transformation 
(\ref{trafo})
with $n = n^k + \Delta n$ generates an extra contribution to the heat 
conduction since 
\begin{eqnarray}
\rho^k V^\mu V^\nu
&=&  \rho^k U^\mu U^\nu - \nonumber \\
& & \frac{\rho^k}{n} \; U^{\mu} J_{\rm{diff}}^{\nu}
-
\frac{\rho^k}{n} \; J_{\rm{diff}}^{\mu} U^\nu
+ {\cal O}(J_{\rm{diff}}^2)
\; .
\end{eqnarray}
It remains to find the relation between the kinetic, (\ref{p2}), and the 
hydrodynamic, (\ref{c2}), definition of the energy density.
The point is that the definitions of temperature in the approach of Eckart 
and in the kinetic theory under consideration are different \cite{Wei}.
In kinetic theory there is a unique way to define temperature as the 
temperature of the leptons and photons which stay in thermal equilibrium
during and after the kinetic decoupling of the neutralinos.
In the approach of Eckart the temperature was chosen
such that the energy density agrees with the one in the adiabatic limit.
Thus it is clear that the difference in the definitions
should be generated by $\Delta \rho$. Since we are only
interested in effects linear in the collision time
we may expand $\Delta \rho$ in a first order Taylor expansion.
Solving this expansion for the difference in the temperatures 
\begin{equation}
T_{\rm{R}}
=
T + 
\left(
\frac{\partial \rho}{\partial T} 
\right)_n^{-1}
\left[
\left(\frac{\partial \rho}{\partial n}\right)_T \Delta n
-\Delta \rho
\right]
\; .
\end{equation}
Let us now rewrite the energy-momentum tensor as calculated
in the kinetic theory in terms of $T$, $n$ and $U$:
\begin{eqnarray}
&&T^{k\; \mu\nu}(T,n)
=
\rho (T,n) U^\mu U^\nu - P(T,n) h^{\mu\nu}
\nonumber \\
&&+
\left[
\left(
\frac{\partial P}{\partial n}
\right)_\rho 
\Delta n
+
\left(
\frac{\partial P}{\partial \rho} 
\right)_n 
\Delta \rho 
-
\Delta P
\right] h^{\mu\nu}
\nonumber \\
&&+ 
T_B^{(1)\; \mu\nu}(T,n) + \frac{w}{n} U^\mu J_{\rm{diff}}^\nu (T,n)
+ \frac{w}{n} J_{\rm{diff}}^\mu(T,n) U^\nu
\nonumber \\
&&+ T_C^{(1)\; \mu\nu}(T,n) \; .
\end{eqnarray}

This expression can be compared to (\ref{emt1}) and the transport coefficients 
can be extracted. We express them in terms of the functions $f^{(i,j)}_a$,
\begin{eqnarray}
\zeta &=&
f_0^{(1,3)} \left(\frac{\partial P}{\partial \rho}\right)_n^2
+
f_0^{(1,1)} \left(\frac{\partial P}{\partial n}\right)_\rho^2
\\ \nonumber \\
&-&\frac{2}{3} f_0^{(3,1)} \left(\frac{\partial P}{\partial \rho}\right)_n
+ \frac{1}{9} f_0^{(5,-1)}
\nonumber \\ \nonumber \\
&+&
2 f_0^{(1,2)} \left(\frac{\partial P}{\partial \rho}\right)_n
\left(\frac{\partial P}{\partial n}\right)_\rho
-\frac{2}{3} f_0^{(3,0)} \left(\frac{\partial P}{\partial n}\right)_\rho 
\; , \nonumber \\ \nonumber \\
\eta
&=&
\frac{1}{15} f_2^{(5,-1)}
\; , \\ \nonumber \\
T\chi
&=&
\frac{1}{3} \left[
f_1^{(3,1)}-2\frac{w}{n} f_1^{(3,0)} + 
\left(\frac{w}{n}\right)^2 f_1^{(3,-1)}
\right]
\; .
\end{eqnarray}
Instead of equating the collision integral
to our expansion (\ref{entw}) and solving
for the unknown functions $\kappa_a(\omega,x)$ $(a\in \{0,1,2\})$
we give a qualitative correct estimate. This can be achieved
by using the following model for the collision integral 
\begin{equation}
L[F^{(1)}]
\approx
-\omega \; \tau^{-1} \; F^{(1)}
\end{equation}
which corresponds to $\kappa_a(\omega,x) = \tau^{-1}$ for all 
$(a\in \{0,1,2\})$. This model reflects the linear dependence
on $F^{(1)}$ and generates the variation of $F^{(0)}$
in the direction of the adiabatic motion through
the rate of elastic scatterings. Furthermore
since neutralinos are non-relativistic
at kinetic decoupling we use the following
approximations
\begin{equation}
f^{(i,j)}_a
\approx 
f^{(i,j)}
\approx
i!! (MT)^{\frac{i-3}{2}} M^j n \tau
\; 
\end{equation}
for odd $i$ and 
\begin{equation}
n 
\approx
\frac{g}{(2\pi)^{3/2}} \; (MT)^{3/2} \; 
{\rm exp} \left(\alpha-\frac{M}{T}\right)
\; .
\end{equation}
In this case $\Delta \rho$ and $\Delta P$ depend linearly on $\Delta n$: 
$\Delta \rho \approx M \Delta n\;, \Delta P \approx 5/3 T \Delta n$. 
At any time step $\Delta t$ the variation of the number density due to 
non-equilibrium processes is given by the number of collisions during 
$\Delta t$, i.~e., $\Delta n=\dot{n} \tau$.  

We find in ${\cal O}(\tau)$ and up to order $T/M$ 
\begin{equation}
\eta \approx nT\tau , \quad
\zeta \approx \frac{5}{3} nT\tau, \quad
\chi \approx 0 \; .
\end{equation}
It is interesting to note that $\chi \approx 0$ 
at this order since the contribution of the transformation (\ref{trafo})
to the energy momentum tensor cancels $T_B^{(1)}$. 

On the first sight it might be surprising that heat conduction vanishes
and bulk viscosity is nonvanishing. The mentioned cancellation between 
$T_B^{(1)}$ and $J_{\rm diff}$ indicates that the only possible mechanism 
to transport heat in the neutralino fluid is convection. Since the 
neutralinos are very slow and very sparse heat can neither be radiated nor 
conducted. We decided to use the frame that is comoving with the neutralinos
(Eckart's approach \cite{Eck}), thus there is no heat conduction here. 
In a single fluid bulk viscosity goes along with internal degrees of freedoms 
or with particle production or decay. In our situation the number of 
neutralinos is conserved and they do not have any internal degrees of 
freedom which can dissipate energy. Nevertheless, the bulk viscosity is 
non-zero, the reason is that we are dealing with two fluids and the bulk 
viscosity describes the energy dissipation to the radiation fluid. There 
should be a corresponding term for the radiation fluid, however we can 
neglect this term since $\rho_{\rm R} \gg nT$ at kinetic decoupling. The 
authors of Ref.~\cite{BFS} have incorrectly assumed that $\chi \neq 0$ 
and $\zeta = 0$ in their work. Let us note that our result ($\chi = 0$ and 
$\zeta \neq 0$) holds in general for any kind of WIMPy CDM.

\section{Collisional damping of acoustic perturbations}

The viscosity coefficients and the coefficient for
heat conduction enter in the decay rate of
acoustic perturbations, which we will study now.
Following Weinberg \cite{Wei} let us start with a static
homogeneous fluid with
\begin{equation}
U=(1,\vec{0})\; , \hspace{0.2cm}
\rho, P, n, T = \rm{const}
\; .
\end{equation}
This fluid should leave the adiabatic limit
but stay close to thermal equilibrium.
As a consequence small perturbations
will occur with the space-time dependence
\begin{eqnarray}
\label{pert}
\delta (\rho, P, n, T, \vec{k} \cdot \vec{U})
&=&
\left(\rho^{(1)}, P^{(1)}, n^{(1)}, T^{(1)}, 
\vec{k} \cdot \vec{U}^{(1)}\right) \nonumber \\ \nonumber \\
&&\rm{exp}\left(i\omega t\right)
\rm{\exp}\left(-i\vec{k}\cdot \vec{x}\right)
\; .
\end{eqnarray}
Note that the perturbation of the zeroth component of $U$
should vanish in order to guarantee the normalization
condition.

Inserting (\ref{pert}) in the conservation laws
for the number density, energy density and momentum
we get a system of three linear algebraic equations
\begin{equation}
\label{sys}
{\cal A}(\zeta,\eta)
\;
\left(\delta T, \delta n, \vec{k}\cdot \delta\vec{U} \right)^{T}
=
\vec{0}
\; ,
\end{equation}
where we used $\chi\equiv 0$ and
the matrix ${\cal A}(\zeta,\eta)$ is given in \cite{Wei}.
The dispersion relation is provided by the requirement
\begin{equation}
\rm{det} {\cal A}(\zeta,\eta)
=
0 \ ,
\end{equation}
which yields to first order in the collision time ($k \equiv |\vec{k}|$) 
\begin{equation}
{\rm Re} \; \omega = k v_s \;, \hspace{0.2cm}
{\rm Im} \; \omega
=
- L[\zeta,\eta] \; k^2 \equiv - \Gamma[\zeta,\eta] 
\;.
\end{equation}
The square of the isentropic sound speed, $v_s$, is given by  
\begin{eqnarray}
v_s^2 &\equiv& \left({\partial P\over \partial \rho}\right)_\sigma \nonumber \\
    &=&  {T\over \rho+P} \left({\partial P\over \partial T}\right)_n 
         \left({\partial P\over \partial \rho}\right)_n 
         + {n\over \rho +P} \left({\partial P\over \partial n}\right)_T 
    \nonumber  \\
    &\approx& \frac 53 {T \over M_{\tilde{\chi}}}
\end{eqnarray}
and the characteristic length $L$ for absorption reads 
\begin{equation}
L[\zeta,\eta]
\approx
\frac{\zeta + \frac{4}{3} \eta}{2\rho}
\approx
\frac{3}{2} \frac{T}{M_{\tilde{\chi}}}\;  \tau
\; .
\end{equation}
Note that the length scale of collisional damping is proportional
to the relaxation time. The authors of \cite{BFS} assume instead
that the characteristic scale for acoustic absorption
is given by the collision time. This is correct only 
if acoustic perturbations are smeared out
after a single contact with the heat bath.
We proved already in Sec.~III that in the case under consideration
a huge number of contacts with the heat bath is needed
to establish equilibrium. 

Since the parameters of the fluid are slowly varying
during the cooling to $T_{\rm kd}$, the amplitude of an acoustic perturbation
behaves like a WKB solution. The damping of density perturbations is given by 
\begin{eqnarray}
\delta(k) 
&=&
\rm{exp}\left(
-\int_0^{t(T_{\rm kd})} \Gamma(t) {\rm d} t \right)
\; , \nonumber \\
&=&
\label{delta}
{\rm exp}\left(-\frac{3}{10} \frac{T_{\rm kd}}{M_{\tilde{\chi}}}
\left(\frac{k_{\rm phys}}{H}\right)^2_{T=T_{\rm kd}} \right)
\; .
\end{eqnarray}
We integrate over the time interval $[0,t_{\rm kd}]$
during which the fluid is close to thermal equilibrium
and CDM density perturbations evolve like (damped) sound waves. In 
principle $t_{\rm kd}$ is a function of $k$, however we find that 
for modes of interest we can take $t_{\rm k}$ to be independent of $k$.  
This follows as ${\rm Re}(\omega) \tau = v_s k_{\rm phys} \tau < 1$ is 
easily fulfilled for the subhorizon scales $(k_{\rm phys}/H)(T_{\rm kd}) 
< 1/v_s \sim 10^4$ for typical MSSM masses, including all modes of 
interest.

{}From (\ref{delta}) we can read off a typical wavelength for collisional 
damping
\begin{equation}
l_{\rm d}(T_{\rm kd}) 
=
\frac{2\pi}{\sqrt{10}} v_{\rm kd} \; R_H(T_{\rm kd})
\; ,
\end{equation}
where $v_{\rm kd} = \sqrt{3 T_{\rm kd}/M_{\tilde{\chi}}}$ and $R_H \equiv
1/H$ denotes the Hubble radius. We find $l_{\rm d} \approx 0.06 (0.05) 
R_H(T_{\rm kd})$ for $M_{\tilde{\chi}} = 100 (150)$ GeV and $M_{\tilde{L}} 
= 200 (250)$ GeV.

\begin{figure}
\resizebox{0.95 \linewidth}{!}
{\includegraphics{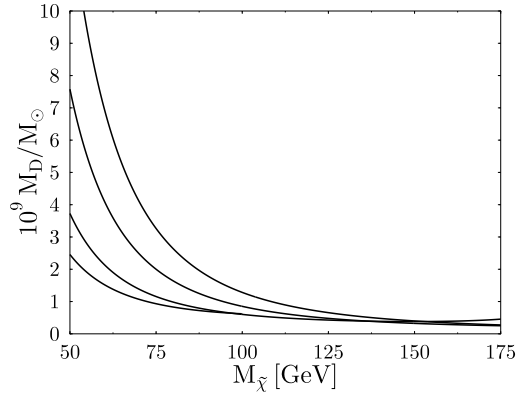}}
\caption{The damping scale $M_{\rm d}$ in solar masses as a function of
the neutralino mass $M_{\tilde{\chi}}$ for different
values of the slepton Mass $M_{\tilde{L}}=150, 200, 300, 400$
GeV. The slepton mass increases from the bottom to the top.
\label{fig4}}
\end{figure}

Instead of characterizing acoustic perturbations by their wavelength or 
wavenumber $k_{\rm phys} \sim a^{-1} \sim t^{\frac{1}{2}}$ it is more 
convenient to work with a constant in time---the rest mass $M$ of  
neutralinos within a sphere of radius $2\pi / k_{\rm phys}$:
\begin{equation}
M \equiv
\frac{4}{3} \pi 
\left(2\pi / k_{\rm phys}\right)^3
\; n_{\tilde{\chi}} M_{\tilde{\chi}}
\; .
\end{equation}
Using the definition of $M$ we can write (\ref{delta}) as
\begin{equation}
\delta(M)
=
{\rm exp}\left[-
\left(M_{\rm d}/M \right)^{\frac{2}{3}}\right]
\; ,
\end{equation}
where the mass scale of damping, $M_{\rm d}$, is given by 
\begin{eqnarray}
M_{\rm d} &=& \frac{2^4 \pi^4}{5} 
\left(\frac{3}{10}\right)^{\frac 12} 
\left(\frac{T_{\rm kd}}{M_{\tilde{\chi}}}\right)^{\frac 32}
M_{\tilde{\chi}} n_{\tilde{\chi}}(T_{\rm kd}) R_H^3(T_{\rm kd}) 
\nonumber \\
&\approx& 2.6 \times 10^{-8} 
\frac{(1\ \mbox{\rm GeV})^3}{(M_{\tilde{\chi}} T_{\rm kd})^{3/2}}
\omega_{\tilde{\chi}} M_\odot \ .
\end{eqnarray}
Exploring the parameter space of the MSSM
we find typically $M_D\approx 10^{-9} M_\odot$, cf. Fig.~\ref{fig4}.
%%%%%%%%%%%%%%%%%%%%%%%%%%%%%%%%%%%%%%%%%%%%%%%%%%%%%%%%%%%%%%%
%%%%%%%%%%%%%%%%%%%%%%%%%%%%%%%%%%%%%%%%%%%%%%%%%%%%%%%%%%%%%%

\section{Free streaming}

\begin{figure}
\resizebox{0.95 \linewidth}{!}
{\includegraphics{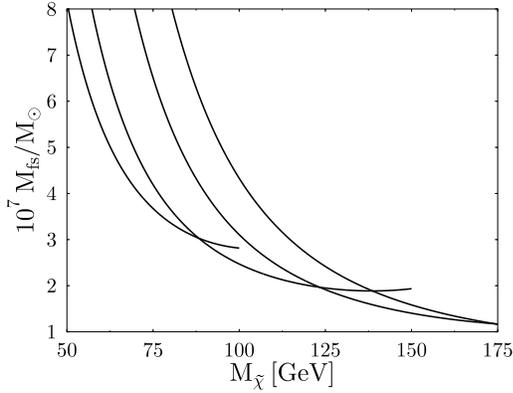}}
\caption{The free streaming mass $M_{\rm fs}$ 
at matter-radiation equality
in solar masses
as a function of the neutralino mass $M_{\tilde{\chi}}$ for different
values of the slepton mass $M_{\tilde{L}}=150, 200, 300, 400$
GeV. The slepton mass increases from the bottom to the top.
\label{fig5}}
\end{figure}

For temperatures $T<T_{\rm ls}$ the neutralinos are collisionless so 
that the viscosity coefficients vanish. Each neutralino moves along a 
geodesic in space-time. This geodesic motion of neutralinos provides a 
second damping mechanism: free streaming \cite{fs}. If the proper distance 
$l_{\rm fs}(t)$ which a neutralino can travel along a geodesic in time $t$ 
is larger then then the proper wavelength $\lambda_{\rm phys}\equiv 
2\pi/k_{\rm phys}$ of a perturbation at $t$ any structure will be wiped out 
since the neutralinos will propagate from an overdense region to an 
underdense region. The proper distance of free streaming for a neutralino 
at time $t\in [t_{\rm ls},t_{\rm eq}]$ is given by
\begin{equation}
l_{\rm fs}(a)
=
\frac{a_{\rm ls}}{a} \; v_{\rm ls} \;
{\rm ln}\left(\frac{a}{a_{\rm ls}}\right)
\; R_H(a) \; ,
\end{equation}
where $a_{\rm ls}$ denotes the expansion factor
and $v_{\rm ls}=\sqrt{3 T_{\rm ls}/M_{\tilde{\chi}}}$
the average neutralino velocity at last scattering.
Exploring the parameter space of the MSSM we find 
at equality typically $l_{\rm fs}(a_{\rm eq})\approx
10^{-8} \; R_H(a_{\rm eq})$, which corresponds to 
$5 \times 10^{-4}$ pc today. This length scale contains a
mass of $M_{\rm fs} \approx 10^{-7} M_{\odot}$, 
cf.~Fig.~\ref{fig5}.

Let us compare the scales of the two distinct damping mechanism at 
equality, the time when structures begin to grow (before equality CDM 
perturbations grow logarithmically). We find for $a\gg a_{\rm ls}$
\begin{equation}
\frac{l_{\rm fs}}{l_{\rm d}}
=
\frac{\sqrt{10}}{2\pi}\; 
{\rm ln}\left(\frac{a}{a_{\rm ls}}\right)
\; ,
\end{equation}
which gives $l_{\rm fs} / l_{\rm d} \approx 6$ 
(or $M_{\rm fs}/M_{\rm d} \approx 220$) at equality. 

%%%%%%%%%%%%%%%%%%%%%%%%%%%%%%%%%%%%%%%%%%%
%%%%%%%%%%%%%%%%%%%%%%%%%%%%%%%%%%%%%%%%%%%%

\section{Conclusions}

We have shown in this paper that kinetic decoupling of neutralino dark
matter leads to collisional damping at the scale $10^{-9} M_\odot$. This scale
could be larger for certain regions in the MSSM parameter space, e.g., when
the neutralino mass and one of the slepton masses (probably the stau) are 
nearly degenerate. In that case our tree-level expressions become singular, 
and are not applicable. We have pointed out that it is important to distinguish
between the collision time and the relaxation time of neutralino CDM. The 
corresponding temperatures differ by about an order of magnitude, which can
lead to a difference of several orders of magnitude in the corresponding
mass scales of damping.

The process of collisional damping has been described by imperfect fluids,
and we calculated the transport coefficients from kinetic theory, by 
generalizing the method of Straumann \cite{Strau} in order to include massive 
particles. We found 
that bulk viscosity can not be neglected, whereas heat conduction is negligible
in the process of kinetic decoupling of neutralinos.

After kinetic decoupling free streaming starts to smear out remaining 
perturbations on scales below $10^{-7} M_\odot$ by the time of equality. Both 
scales are quite close, which shows that both mechanisms have to be 
considered in the calculation of the resulting power spectra for cold dark 
matter. We will present the corresponding transfer functions and power spectra
elsewhere \cite{HS2}.
 
These damping mechanisms provide a sharp (exponential) cut-off in the power 
spectrum of CDM objects. Such a cut-off sets the scale for the very first
objects that form in hierarchical structure formation. Although this might be
impossible to observe directly, it might have implications on the 
substructure of galactic halos and on the structure of CDM in void regions, 
where some of the first CDM clouds might have a chance to survive. The 
cosmological and astrophysical consequences of this cut-off will be 
investigated in a forthcoming publication \cite{HS2}.

\section*{Acknowledgement}

We acknowledge discussions with H.~Eberl and V.C.~Spanos.
D.J.S. thanks the Alexander-von-Humboldt foundation and the 
Austrian Academy of Sciences for financial support.

\begin{appendix}

\section{Cross sections}

This appendix contains the exact scattering amplitudes
at tree level for elastic scattering 
and annihilation processes and simplified formulas
for the related cross sections.

Let us begin with the squared transition matrix element
$\mid{\cal T}\mid^2$ (summed over final and averaged
over initial spins) for
$\tilde{B} + \{L,\bar{L}\} \rightarrow \tilde{B} + \{L,\bar{L}\}$
expressed as a function of the usual Mandelstam variables.
Our notation is as follows: ${\cal T}_l(u)$ is
the scattering amplitude which describes the exchange
of a left-handed slepton in the $u$ channel and so on.
For the squared terms we find
\begin{eqnarray}
\mid{\cal T}_l(u)\mid^2
&=&
4 \left(g b_L\right)^4 
\left(
\frac{u-M_{\tilde{B}}^{\; \; 2}-m_L^{\; \; 2}}
     {u-M_{\tilde{L}}^{\; \; 2}}
\right)^2
\nonumber \; ,\\
\mid{\cal T}_l(s)\mid^2
&=&
4 \left(g b_L\right)^4 
\left(
\frac{s-M_{\tilde{B}}^{\; \; 2}-m_L^{\; \; 2}}
     {s-M_{\tilde{L}}^{\; \; 2}}
\right)^2
\nonumber \; ,\\
\mid{\cal T}_r(u)\mid^2
&=&
4 \left(g c_L\right)^4 
\left(
\frac{u-M_{\tilde{B}}^{\; \; 2}-m_L^{\; \; 2}}
     {u-M_{\tilde{L}}^{\; \; 2}}
\right)^2
\nonumber \; ,\\
\mid{\cal T}_r(s)\mid^2
&=&
4 \left(g c_L\right)^4 
\left(
\frac{s-M_{\tilde{B}}^{\; \; 2}-m_L^{\; \; 2}}
     {s-M_{\tilde{L}}^{\; \; 2}}
\right)^2
\nonumber \; ,
\end{eqnarray}
where we assumed for simplicity 
that the masses of the left- and
right-handed sleptons are equal.

For the different interference terms
we find
\begin{eqnarray}
{\cal T}_l(u) {\cal T}_l^{\; \dagger}(s)
&=&
4 \left(g b_L\right)^4
\frac{2m_L^2 M_{\tilde{B}}^{\; \; 2} 
- M_{\tilde{B}}^{\; \; 2} t} {\left(u-M_{\tilde{L}}^{\; \; 2}\right)
\left(s-M_{\tilde{L}}^{\; \; 2}\right)} 
\nonumber \; ,\\
{\cal T}_r(u) {\cal T}_r^{\; \dagger}(s)
&=&
4 \left(g c_L\right)^4
\frac{2m_L^2 M_{\tilde{B}}^{\; \; 2} 
- M_{\tilde{B}}^{\; \; 2} t}
     {\left(u-M_{\tilde{L}}^{\; \; 2}\right)
      \left(s-M_{\tilde{L}}^{\; \; 2}\right)}
\nonumber \; ,\\
{\cal T}_l(u) {\cal T}_r^{\; \dagger}(u)
&=&
16 g^4 \left(b_L c_L\right)^2
\left(\frac{m_L M_{\tilde{B}}}
     {u-M_{\tilde{L}}^{\; \; 2}}\right)^2
\nonumber \; ,\\
{\cal T}_l(s) {\cal T}_r^{\; \dagger}(s)
&=&
16 g^4 \left(b_L c_L\right)^2
\left(\frac{m_L M_{\tilde{B}}}
     {s-M_{\tilde{L}}^{\; \; 2}}\right)^2
\nonumber \; ,\\
{\cal T}_l(u) {\cal T}_r^{\; \dagger}(s)
&=&
4 g^4 \left(b_L c_L\right)^2
\frac{2m_L^{\; \; 2} M_{\tilde{B}}^{\; \; 2} 
- m_L^{\; \; 2} t}
     {\left(u-M_{\tilde{L}}^{\; \; 2}\right)
      \left(s-M_{\tilde{L}}^{\; \; 2}\right)}
\; .\nonumber
\end{eqnarray}
Summing up the squared and interference terms yields
\begin{eqnarray}
\mid {\cal T} \mid^2
&=&
4 g^4 
\left(b_L^{\; \; 4} + c_L^{\; \; 4} \right)
\nonumber \\
&&\left[
\left(
\frac{
s-M_{\tilde{B}}^{\; \; 2} - m_L^{\; \; 2}
}{
s - M_{\tilde{l}}^2
}
\right)^2
+ \left(
\frac{
u-M_{\tilde{B}}^{\; \; 2} - m_L^{\; \; 2}
}{
u - M_{\tilde{l}}^2
}
\right)^2
\right. \nonumber \\
&&\left. \hspace{0.5cm} + 
\frac{
\left(2m_L M_{\tilde{B}}\right)^2 - 2M_{\tilde{B}}^{\; \; 2} \; t
}{
\left(s-M_{\tilde{L}}^{\; \; 2}\right) 
\left(u-M_{\tilde{L}}^{\; \; 2}\right)
}
\right]
\nonumber \\
&+&8g^4\left(b_L c_L \right)^2
\nonumber \\
&&\left[
\left(
\frac{
2m_L M_{\tilde{B}}
}{
s-M_{\tilde{L}}^{\; \; 2}
}
\right)^2
+
\left(
\frac{
2m_L M_{\tilde{B}}
}{
u-M_{\tilde{L}}^{\; \; 2}
}
\right)^2
\right.
\nonumber \\
&&\left. \hspace{0.5cm} +
\frac{
\left(2m_L M_{\tilde{B}}\right)^2 - 2m_{L}^{\; \; 2} \; t
}{
\left(s-M_{\tilde{L}}^{\; \; 2}\right) 
\left(u-M_{\tilde{L}}^{\; \; 2}\right)
}
\right]
\; . \nonumber
\end{eqnarray}
Since $m_L$ denotes the mass of a standard model fermion
we use $m_L\approx 0$. Furthermore the binos are
nonrelativistic so that 
$s\approx M_{\tilde{B}}^{\; \; 2} + 2M_{\tilde{B}} E_L$
to a very good approximation. Using the mentioned
simplifications we find
\begin{eqnarray}
\mid {\cal T} \mid^2
&=&
256 \left(b_L^4+c_L^4\right)
\left(
\frac{G_{\rm F} M_W^2}
{M_{\tilde{L}}^{\; \; 2}-M_{\tilde{B}}^{\; \; 2}}
\right)^2
\left(1-\frac{t}{4E_L^{\; \; 2}} \right)
\nonumber
\end{eqnarray}
and for the elastic cross section the simplified formula
\begin{eqnarray}
\sigma_{\rm el}(E_L)
=
\frac{24}{\pi} \left(b_L^4+c_L^4\right)
\left(
\frac{G_{\rm F} M_W^2}
{M_{\tilde{L}}^{\; \; 2}-M_{\tilde{B}}^{\; \; 2}}
\right)^2
\; E_L^{\; \; 2}
\; .\nonumber
\end{eqnarray}

The squared transition matrix element (summed over final
and averaged over initial spins)
for ${\tilde B} + {\tilde B} \rightarrow \bar{F} + F$ 
may be found from $\mid {\cal T} \mid^2$
for elastic scattering processes
by making the following modifications:
$s\rightarrow u$, $t\rightarrow s$ and $u\rightarrow t$.
Expanding in $m_F/M_{\tilde{B}}$ and in 
the Lorentz invariant relative velocity $v$
up to second order yields
\begin{eqnarray}
v\sigma_{\rm ann}
&=&
\frac{2}{\pi}
\left(
\frac{G_{\rm F} M_W^2}
{M_{\tilde{F}}^{\; \; 2}+M_{\tilde{B}}^{\; \; 2}}
\right)^2
\left[
\left(b_F^{\; \; 2} + c_F^{\; \; 2}\right)^2 m_F^{\; \; 2}
\right. \nonumber \\
&&+
\left.\frac{2}{3} \left(b_F^{\; \; 4} + c_F^{\; \; 4}\right)
\frac{
M_{\tilde{F}}^{\; \; 4}+M_{\tilde{B}}^{\; \; 4}
}{
\left(
M_{\tilde{F}}^{\; \; 2}+M_{\tilde{B}}^{\; \; 2}
\right)^2
}
\left(M_{\tilde{B}} v\right)^2
\right]
\nonumber \; ,
\end{eqnarray}
where $v$ is given by $(v/2)^2 = 1-(2M_{\tilde{B}})^2/s$.

\section{Adiabatic relations}

In this appendix we show how $\dot{T}$ and space-time gradients of $\alpha$
are related to $U^\lambda_{\; ,\lambda}$ and $Q_\mu$.

The second law of thermodynamics gives the variation in the entropy per 
particle $\sigma$ as
\begin{equation}
\label{tod}
n T {\rm d}\sigma
=
{\rm d}\rho - \frac{w}{n}\; {\rm d}n
\; .
\end{equation}
Since ${\rm d}\sigma$ must be a perfect differential  
\begin{equation}
\label{per}
T \; \left(\frac{\partial P}{\partial T}\right)_n
=
w-n \; \left(\frac{\partial\rho}{\partial n}\right)_T
\end{equation}
follows. For adiabatic motion 
\begin{eqnarray}
0 &=& n T \dot{\sigma} \\ \nonumber \\
&=&
\left(\frac{\partial\rho}{\partial T}\right)_n \; \dot{T}
- \frac{T}{n} \; 
\left(\frac{\partial P}{\partial T}\right)_n \; \dot{n}
\; ,\nonumber
\end{eqnarray}
or, using the conservation law ${\rm d}(n U)=0$
\begin{equation}
\dot{T}
\label{dT}
=
-T \; \left(\frac{\partial P}{\partial \rho}\right)_n
\; U^\lambda_{\; ,\lambda}
\; .
\end{equation}
The Gibbs-Duhem relation gives the variation
of the pseudochemical potential $\alpha=\mu/T$ as
\begin{equation}
\label{GDR}
{\rm d}\alpha
=
\frac{{\rm d}P}{nT} - \frac{w}{nT} \; \frac{{\rm d}T}{T}
\; .
\end{equation}
For adiabatic motion and using (\ref{per}) and (\ref{dT})
this yields
\begin{equation}
T \; \dot{\alpha}
=
- \left(\frac{\partial P}{\partial n}\right)_\rho \; 
U^\lambda_{\; ,\lambda}
\; .
\end{equation}
Using (\ref{GDR}) and the relativistic generalization
of the Euler equation we find 
\begin{equation}
T \; h^{\mu\lambda} \alpha_{\; ,\lambda}
=
- \frac{w}{Tn} \; h^{\mu\lambda} \; Q_\lambda
\; .
\end{equation}
Thus the variation of $\alpha$ in the plane perpendicular
to the adiabatic flow is generated by the projection
of the heat current on this plane.
\end{appendix}


\begin{thebibliography}{MM8888}
\bibitem{ss} The scale invariance of hierarchical structure formation 
   obviously
   has to be broken at some small scale, otherwise the contribution of 
   density fluctuations to the local energy density would be ultra-violet 
   singular.
\end{thebibliography}
\end{document}